\documentclass[preprint2]{aastex631}
\begin{document}
\journalinfo{To be accepted for publication}
\shorttitle{The Cepheid hosting binary cluster: NGC 7790 and Berkeley 58}
\title{A suite of classical Cepheids tied to the binary cluster Berkeley 58 \& NGC 7790}

\author{Daniel Majaess}
\affiliation{Mount Saint Vincent University, Halifax, Canada}
\email{Daniel.Majaess@msvu.ca}

\author{David G.~Turner}
\affiliation{Saint Mary's University, Halifax, Canada}

\begin{abstract}
The classical Cepheids CE Cas A, CE Cas B, CF Cas, and CG Cas are likely members of the binary open cluster comprising NGC 7790 and Berkeley 58.  The clusters are of comparable age and in close proximity, as deduced from differentially dereddened $UuB_PBVGR_P$ photometry, and Cepheid period-age relations.  Gaia DR3 astrometric and spectroscopic solutions for the clusters are likewise consistent. Conversely, the seemingly adjacent open cluster NGC 7788 is substantially younger and less distant.
\end{abstract}


\section{Introduction}

\begin{figure}[t]
\begin{center}
 \includegraphics[width=3.5in]{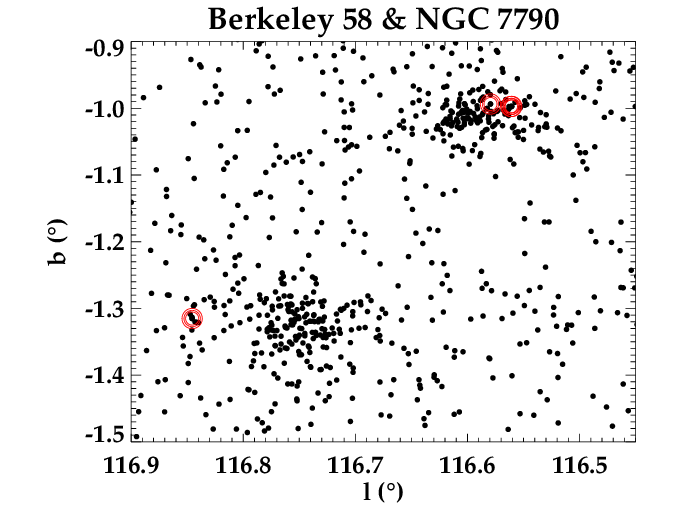} 
  \caption{A subsample of stars ($\pi <0.37$ mas) along the sightline to Berkeley 58 and NGC 7790 (right). Cepheids are encompassed by red circles.}
 \label{figfov}
\end{center}
\end{figure}

\begin{figure*}[t]
\begin{center}
 \includegraphics[width=3.5in]{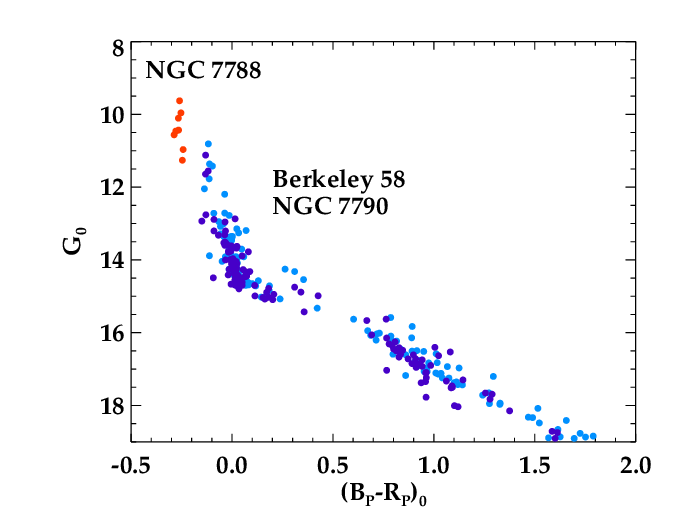} 
 \includegraphics[width=3.5in]{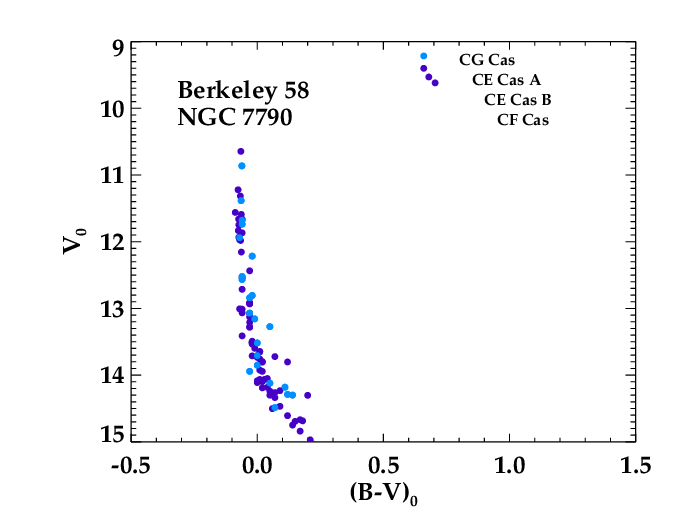} 
  \caption{Differentially dereddened color-magnitude diagrams for Berkeley 58 (blue), NGC 7790 (magenta), and NGC 7788 (red, earlier turnoff). Sequences for Berkeley 58 and NGC 7790 align owing to a similar age and binarity. Left, NGC 7788 is an unassociated younger cluster in the foreground. Right, the Cepheids were dereddened using the \citet{tur16} compilation.}  
 \label{figcmd}
\end{center}
\end{figure*}

The coauthor (D.G.T.) long suspected that the open clusters Berkeley 58 and NGC 7790 (Fig.~\ref{figfov}) might be associated \citep[see also][]{pp21}.  \citet{tu08} established an age for Berkeley 58 of $\log{\tau}=8.0 \pm 0.1$ using \citet{mmm93} isochrones, and \citet{maj13} concluded that NGC 7790 appears the same age according to Padova models \citep[][]{gir02}.   Importantly, the classical Cepheids CE Cas A, CE Cas B, and CF Cas are members of NGC 7790 \citep[][and discussion therein regarding O.~Eggen]{san60}, and \citet{tu08} argued that Berkeley 58 hosts CG Cas as a coronal member (Fig.~\ref{figfov}, near bottom left).

The other binary open cluster that may host a Cepheid is NGC 6716 and Collinder 394 \citep{tp85}.  However, there are ambiguities concerning the membership of the classical Cepheid BB Sgr therein, and a separate effort is underway to clarify that star's status.  

Here, differentially dereddened $UuB_PBVGR_P$ photometry is employed to assess whether Berkeley 58 and NGC 7790 form a binary pair, in conjunction with Gaia DR3 astrometry and spectroscopy.

\section{Analysis}
Gaia DR3 data were utilized to inspect the field of view (Fig.~\ref{figfov}), and the clusters share comparable proper motions (Table~\ref{table:summary}).  Yet  unrelated open clusters along an adjacent sightline (e.g., NGC 7788) feature similar astrometry. Consequently, a holistic approach was pursued whereby cluster ages and potential binarity were examined via dereddened color-magnitude diagrams, along with period-age relations for the Cepheids.  In addition, a debate continues regarding the Gaia zero-point \citep[e.g.,][]{ow22}, and a similar situation transpired for Hipparcos parallaxes \citep[e.g., the Pleiades and Blanco 1,][]{maj11hip}.  Hence the present reliance on dereddened color-magnitude diagrams.

A color-magnitude diagram of differentially dereddened Gaia $B_P G R_P$ photometry is plotted in Fig.~\ref{figcmd}. Stars were individually dereddened using extinction estimates inferred from low-resolution Gaia spectroscopy ($\lambda\simeq330-1050$ nm). \citet{gaia23} and \citet{an23} provide preliminary estimates for $T_{\rm eff}$, $\log{g}$, $A_G$, and $E(B_P-R_P)$. \citet{an23} stress that work on refining their initial approach continues.  Indeed, there are discernible offsets between DR3 spectroscopically dereddened main-sequences and unobscured clusters (e.g., NGC 2451). Nevertheless, Fig.~\ref{figcmd} confirms that Berkeley 58 and NGC 7790 are coeval, whereas NGC 7788 appears younger \citep[see also][]{dav12}. Only the brightest turnoff stars for NGC 7788 are shown in Fig.~\ref{figcmd}, since its main-sequence bisects the older and more distant binary cluster.  Regarding the latter, Berkeley 58 appears \textit{marginally} closer than NGC 7790.  A small subset of rogue points were removed from Fig.~\ref{figcmd}.

\begin{deluxetable*}{lcccccc|c}
\tablecaption{Gaia DR3 and Isochrone results.\label{table:summary}}
\tablehead{\colhead{} & \colhead{Berkeley 58\tablenotemark{ }} & \colhead{NGC 7790} & \colhead{CE Cas A} & \colhead{CE Cas B} & \colhead{CF Cas} & \colhead{CG Cas} & \colhead{ NGC 7788}}
\startdata
$\pi$ & $0.30\pm0.04$ & $0.29\pm0.04$ & $0.31\pm0.02$  & $0.31\pm0.02$ & $0.29\pm0.01$ & $0.27\pm0.01$ & $0.33\pm0.04$ \\
$\mu_{\alpha}$ & $-3.49\pm0.13$ & $-3.24\pm0.10$ & $-3.30\pm0.01$ & $-3.30\pm0.02$ & $-3.24\pm0.01$ & $-3.24\pm0.01$ & $-3.16\pm0.20$ \\
$\mu_{\delta}$ & $-1.81\pm0.11$ & $-1.73\pm0.09$ & $-1.81\pm0.02$ & $-1.87\pm0.02$  & $-1.77\pm0.01$  & $-1.67\pm0.02$ & $-1.80\pm0.11$  \\
\hline
$\log{\tau}$ & $8.0\pm0.1$\tablenotemark{a} & $8.0\pm0.1$\tablenotemark{b} & $7.99\pm0.07$ & $8.03\pm0.06$ & $8.01\pm0.06$ & $8.04\pm0.06$ & $7.3-7.6$\tablenotemark{c}\\
\enddata
\tablenotetext{ }{* Uncertainties for cluster astrometry and Cepheid $\log{\tau}$ represent the standard deviation.}
\tablenotetext{a}{ \citet{tu08}}
\tablenotetext{b}{ \citet{maj13}}
\tablenotetext{c}{ \citet{dav12}}
\end{deluxetable*}

A differential extinction analysis was likewise undertaken using independent $UuBV$ photometry, and the Gaia DR3 astrometric solutions (Table~\ref{table:summary}).  The ultraviolet data utilized are characterized as approximating Johnson $U$ and UVEX Sloan $u$ \citep{mon20}. Standardizing terrestrial ultraviolet photometry is a longstanding challenge \citep[][their \S 8.6]{mon20}, and there exists the Hyades anomaly \citep[e.g.,][]{tu79,maj11hip}.  UVEX $u$ advantageously samples faint stars in both fields (Berkeley 58 \& NGC 7790). Therefore, UVEX $u$ was paired with $BV$ data from \citet{ste00}\footnote{see \citet{pan22}} and \citet{tu08}, and standardized to the coauthor's (D.G.T.) unpublished photoelectric $U$ observations hosted at WebDA\footnote{\url{https://webda.physics.muni.cz/}} \citep{mp03}.  Importantly, the independent $UuBV$ and $B_P G R_P$ results converge upon the same conclusion (Fig.~\ref{figcmd}): Berkeley 58 and NGC 7790 are two clusters of comparable age which are in close proximity, and thus form a binary cluster.  Early-type stars yield mean reddenings of $E(B-V)\simeq0.7$ and $0.5$ for Berkeley 58 and NGC 7790, accordingly, which agree with a subset of published findings \citep{tak88,tu08,maj13}.  Intrinsic $UBV$ colors stemmed from \citet[][and references therein]{tu89}.   The following relationship was adopted to determine the reddening trend, $E(U-B)\simeq E(B-V)X + E(B-V)^2Y + Z$, and constrain remaining photometric inhomogeneities rather than dust properties. A cutoff was imposed for faint Berkeley 58 photometry \citep[e.g., photographic photometry possess large uncertainties,][]{tu08}.

The differential dereddening results were further validated by constructing a $V-BV$ color-magnitude diagram (not shown) tied to the \textit{mean} extinction.  \citet{maj13} determined $<E(B-V)>=0.52\pm0.05$ for NGC 7790 \citep[see also][]{mm88,tak88}, while Berkeley 58 is observed through increased obscuration \citep[i.e., $<E(B-V)>\simeq0.70$,][]{tu08}.  The cluster sequences once again align. 

Cepheid ages can be compared to the clusters using the framework of \citet{bo05}, \citet{tu12}, and \citet{and16}. Pulsation periods for the Cepheids CG Cas ($P\simeq 4^{\rm d}.4$), CF Cas ($P\simeq 4^{\rm d}.8$), CE Cas A ($P\simeq 5^{\rm d}.1$), and CE Cas B ($P\simeq 4^{\rm d}.5$) are comparable. The mean Cepheid ages and standard deviations are $\log{\tau}=8.04\pm0.06, 8.01\pm0.06, 7.99\pm0.07, 8.03\pm0.06$, respectively (Table~\ref{table:summary}).  That matches the evolutionary age of the clusters Berkeley 58 and NGC 7790 \citep[$\log{\tau}=8.0\pm0.1$,][]{tak88,tu08,maj13}.

\section{Conclusions}
A multifaceted approach indicates that Berkeley 58 and NGC 7790 are in close proximity, share a common age, and constitute a binary open cluster (Fig.~\ref{figcmd}, Table~\ref{table:summary}).  That finding is supported by dereddened multiband $UuB_PBVGR_P$ photometry, and DR3 astrometry and spectroscopy.  A suite of four Cepheid members have ages consistent with that for the clusters (i.e., $\log{\tau} \simeq 8.0$, Table~\ref{table:summary}). NGC 7788 is discernibly younger, and lies to the foreground, and is likely unrelated.

Continued research on Cepheid variables in open clusters is desirable \citep[e.g.,][]{and12,che15,bre20,hao22}.

\begin{acknowledgments}
\small{\textbf{Acknowledgments}: this research relies on initiatives such as CDS, NASA ADS, arXiv, Gaia, WebDA (Paunzen, Stütz, Janik, Mermilliod), \citet{ste00}, UVEX.  Janet Drew kindly responded to questions regarding the latter.}
\end{acknowledgments}

\bibliography{article}{}
\bibliographystyle{aasjournal}
\end{document}